\documentclass[a4paper]{article}

\usepackage[dvips]{graphicx, color}

\usepackage{natbib}
\usepackage{amssymb,amsmath}
\usepackage[auth-sc]{authblk}
\usepackage{tikz}
\usetikzlibrary{shapes.geometric, arrows}
\usepackage{enumerate}
\usepackage{epstopdf}
\usepackage{float}
\def\prodint{\mathop{\ensuremath{\lower0.4ex\hbox{\huge $\mathit{\pi}$}}}}
\usepackage{booktabs}

\newcommand{\dif}{{\rm \;d}}
\newcommand{\eins}{{\bf 1}}
\newcommand{\erw}{{\rm E}}

\newcommand{\LM}{{\rm LM}}
\newtheorem{theorem}{Theorem}[section]
\newenvironment{proof}[1][Proof]{\begin{trivlist}
\item[\hskip \labelsep {\bfseries #1}]}{\end{trivlist}}
\providecommand{\keywords}[1]{\textbf{Keywords ---} #1}

\begin{document}

\title{Estimating state occupation and transition probabilities in non-Markov
  multi-state models subject to both random left-truncation and right-censoring}
\author[1]{Alexandra Nie{\ss}l}
\author[1]{Arthur Allignol}
\author[1,2]{Carina M{\"u}ller}
\author[1]{Jan Beyersmann}

\affil[1]{Institute of Statistics, Ulm University, Helmholtzstra{\ss}e 20,
  89069 Ulm, Germany} \affil[2]{Metronomia Clinical Research
  GmbH, Paul-Gerhardt-Allee 42, 81245 M{\"u}nchen, Germany}

\maketitle

\begin{abstract}
  The Aalen-Johansen estimator generalizes the Kaplan-Meier estimator for
  independently left-truncated and right-censored survival data to estimating
  the transition probability matrix of a time-inhomogeneous Markov model with
  finite state space. Such multi-state models have a wide range of
  applications for modelling complex courses of a disease over the course of
  time, but the Markov assumption may often be in doubt. If censoring is
  entirely unrelated to the multi-state data, it has been noted that the
  Aalen-Johansen estimator, standardized by the initial empirical distribution
  of the multi-state model, still consistently estimates the state occupation
  probabilities. Recently, this result has been extended to transition
  probabilities using landmarking, which is, inter alia, useful for dynamic
  prediction. We complement these results in {\color{black} three}
  ways. {\color{black}Firstly,} delayed study entry is a common phenomenon in
  observational studies, and we extend the earlier results to multi-state data
  also subject to left-truncation. Secondly, we present a rigorous proof of
  consistency of the Aalen-Johansen estimator for state occupation
  probabilities, on which also correctness of the landmarking approach hinges,
  correcting, simplifying and extending the earlier result. {\color{black}Thirdly, our   rigorous proof motivates wild bootstrap resampling.} Our developments {\color{black}for left-truncation} are
  motivated by a prospective observational study on the occurrence and the
  impact of a multi-resistant infectious organism in patients undergoing
  surgery. Both the real data example and simulation studies are
  presented. {\color{black}Studying wild bootstrap is motivated by the fact
    that, unlike drawing with replacement from the data, it is desirable to
    have a technique that works both with non-Markov models subject to random
    left-truncation and right-censoring and with Markov models where
    left-truncation and right-censoring need not be entirely random. The
    latter is illustrated for event-driven trials.}
\end{abstract}

\keywords{Nelson-Aalen estimator, Wild bootstrap, Hospital epidemiology, Partly
  conditional transition rate, Methicillin-resistant {\em staphylococcus
   aureus}}

\section{Introduction}\label{sec:intro}

\cite{Aale:Joha:an:1978} developed an estimator of the
transition probability matrix of a non-homogeneous Markov multi-state model
for independently left-truncated and right-censored data. These
  models are useful for studying complex courses of a disease over the course
  of time and applications in medical research include oncology
  \citep{schmoor12:_compet_multis}, cardiology \citep{francesca17:_multi},
  Gastroenterology \citep{jepsen2015clinical}, orthopaedics
  \citep{gillam2012multi} or hospital epidemiology
  \citep{munoz2016handling}. However, the Markov assumption may regularly be
  in doubt in applications. Our motivating data example
  \citep{de2011multistate} investigated the occurrence and the impact of
  Methicillin-resistant {\em staphylococcus aureus} (MRSA) infection in
  hospital compared to patients only colonized with MRSA, using an
  illness-death multistate model. Violations of the Markov assumption arise if
  the time of MRSA infection {\color{black}affects} the hazard of end of hospital stay. See
also Andersen et al. \citep{Ande:Keid:mult:2002} for a clear practical
discussion of a non-Markov multi-state model.

The Markov assumption enters the technical developments in 
\cite{Aale:Joha:an:1978} (see also \cite{ABGK}) in that it implies a
particularly handy form of the intensities of the counting processes of
observed transitions between any two states of the model. Save for the at-risk
processes, these intensities are non-random and equal the usual transition
hazards. This is in contrast to the non-Markov case where the intensities will
also be random through dependence on the past. For instance, in
  the MRSA data, such a dependence is present if the time of infection {\color{black}affects}
  the end-of-stay hazard of an infected patient.

  For non-Markov models and complete observations, Andersen et
  al. \citep[][Section~IV.4.1.4]{ABGK} showed that the Aalen-Johansen
  estimator, standardized by the multinomial estimator of the initial
  distribution of the multi-state model, results in the usual multinomial
  estimator of the unconditional state occupation probabilities. Later,
  \cite{Datt:Satt:vali:2001} observed that this approach still yields a
  consistent estimator of the state occupation probabilities based on
  right-censored observations provided that censoring is entirely
  random. Recently, \cite{putter2016non} extended this approach to a landmark
  Aalen-Johansen estimator of the transition probabilities in randomly
  right-censored non-Markov multi-state models. Their estimator is based on
  Aalen-Johansen estimates of the state occupation probabilities computed on
  subsamples of the data. Consistency of the landmark Aalen-Johansen estimator
  then follows provided that the Aalen-Johansen estimator of the state
  occupation probabilities is consistent.

  However, these findings do not apply to our data example for two reasons:
  Firstly, right-censoring is not much of an issue in hospital epidemiology,
  but delayed study entry, i.e., left-truncation may very well be
  \citep{jb:fogr:2011}. {\color{black}Remarkably, left-truncation was already
    contained in the seminal paper by \cite{km}, see their
    Section~2.} Our example considers a prospective cohort of patients
  colonized with MRSA. The time scale of interest is time since hospital
  admission, and patients may have a delayed study entry if a positive
  laboratory result is only available some time after
  admission. Left-truncation is a common phenomenon in observational studies
  \citep{bluhmki2017time} and an extension of the findings mentioned above
  would be generally useful.

  Secondly, the arguments of \cite{Datt:Satt:vali:2001} are compromised by
  their use of non-random intensities, which do \emph{not} apply in non-Markov
  models{\color{black}, see \cite{cm15} and \cite{overgaard2019}}. This also
  {\color{black}affects} the recent extension of \cite{putter2016non}, because
  their arguments hinge on consistency of state occupation probability
  estimation in the landmark data sets. The issue is this: Datta and Satten
  built on the result of \cite{ABGK} for complete data. Their idea was to show
  that the multivariate Nelson-Aalen estimator, of which the Aalen-Johansen
  estimator is a transformation, consistently estimates the same limit both in
  the completely observed and in the randomly right-censored case. Consistency
  of estimating the state occupation probabilities via the Aalen-Johansen
  estimator then follows from the continuous mapping theorem, although this
  approach was not taken by \cite{Datt:Satt:vali:2001}. For the multivariate
  Nelson-Aalen estimator, \cite{Datt:Satt:vali:2001} started with complete
  data, used martingale methods similar to \cite{Aale:Joha:an:1978} for the
  Markov case and then transferred results to the right-censored case using
  inverse probability of censoring weights
  \citep{horvitz1952generalization}. However, their arguments used intensities
  that were, save for the at-risk processes, non-random (see their Equation
  (A.5)), and use of inverse probability of censoring weights makes the
  arguments unnecessarily complicated. {\color{black} \cite{overgaard2019} took
    a different approach and showed the consistency of the Aalen-Johansen
    estimator for state occupation probabilities based on interval functions
    without using martingale arguments.} {\color{black}Our approach differs from
    Overgaard's in that we will use martingale methods, working, however, with
    the proper intensities. This approach allows to incorporate
    left-truncation and lends itself to wild bootstrap resampling which we
    will find preferable to simple drawing with replacement from the data.}

  The {\color{black}main} aim of this paper is to establish consistent
  estimation of both state occupation and transition probabilities in
  non-Markov models subject to both random left-truncation and
  right-censoring. The main technical results are in
  Section~\ref{sec:main}. {\color{black}Here, we start by considering the
    Nelson-Aalen estimator as an estimator of cumulative partly conditional
    transition rates, which differ from the transition intensities by only
    conditioning on the immediate past, but not on the entire history.}
  Simulations are in Section~\ref{sec:simu}. {\color{black} Here, we assess the
    performance of the Aalen-Johansen estimator for the state occupation
    probabilities and the landmark Aalen-Johansen estimator for the transition
    probabilities in non-Markov models. Additionally, we report results on two
    different resampling methods to obtain confidence intervals. Firstly, we
    profit from the i.i.d.\ data structure under random left-truncation and
    right-censoring, which allow us the use of Efron's bootstrap. Secondly, we
    exploit our result on the consistency of cumulative transition hazards
    estimation in non-Markov models to apply the more flexible wild bootstrap
    resampling technique \citep{bluhmkiWB}. As the wild bootstrap approach
    does not {\color{black}necessarily} require an i.i.d.\ data structure we
    evaluate its performance also in a Markov model subject to
    {\color{black}event-driven} type II censoring.}  An analysis of the MRSA
  data is in Section~\ref{sec:data} and a discussion is in
  Section~\ref{sec:disc}. Proofs are in the appendix; here, we improve on the
  arguments of \cite{Datt:Satt:vali:2001} by working with the proper
  intensities which, in a non-Markov model, will also be random through
  dependence on the past {\color{black}(see Equation~\eqref{eq:5komma5}
    below)}. We will directly work with the observed counting processes such
  that data may also be randomly left-truncated and inverse probability
  weighting is not needed.

\section{Main technical results}\label{sec:main}
Let~$(X(t))_{t\ge 0}$ be a stochastic process with state space~$\{0,1, \ldots,
K\}$. This multi-state process may be non-Markov {\color{black}with a
  non-degenerate initial distribution}. The first aim is to estimate the state
occupation probabilities
\begin{equation}
  \label{eq:1}
  P(X(t)=m),\, m\in \{0,1, \ldots,
  K\}, t\in [0,\tau].
\end{equation}
In a second step, we will also estimate transition probabilities
  \begin{equation}
    \label{eq:0}
    P_{lm}(s,t) =  P(X(t)=m\, |\, X(s)=l),\, l,m\in \{0,1, \ldots,
  K\}, s\le t\in [0,\tau],
  \end{equation}
  using an estimator of $P(X(t)=m)$ in a landmark (sub-) data set that
  accounts for conditioning on $X(s)=l$. Landmarking for such a purpose has
  been proposed by \cite{lida} for the special case of an
  illness-death model and later, for general multistate models, by \cite{putter2016non}.
To this end, define the partly conditional~$l\to m$ transition rate
\citep{Pepe:Cai:some:1993}
\begin{equation}
  \label{eq:3}
  \alpha_{lm}(t) = \lim_{\Delta t \searrow 0} P(X(t + \Delta
  t)=m\,|\,X(t)=l),\, l,m \in \{0,1, \ldots,
  K\}, l\neq m,
\end{equation}
with cumulative quantities~$A_{lm}(t) = \int_0^t \alpha_{lm}(u) \dif u$. 

We assume that observation of~$X$ is subject to random left-truncation by~$L$
and right-censoring by~$C$. Denote the event of study entry, i.e., X reaches
an absorbing state after~$L$, by~$Z$. Given study entry, consider i.i.d.\ data
$(X_i(t))_{t\in (L_i, C_i\wedge T_i]}$, $i=1,\ldots,n$, of $n$ individuals
under study, where~$T_i$ is ~$i$'s time until absorption and~$\wedge$ denotes
the minimum. Let $\mathcal{G}(t)$ denote the self-exciting filtration of the
observed data $(X_i(t))_{t\in (L_i, C_i\wedge T_i]}$, $i=1,\ldots,n$.

Define the individual counting process
\begin{equation}
  \label{eq:4}
    N_{i;lm}(t) = \#\mbox{\ of observed $l\to m$ transitions of $i$ in $[0,t]$},
\end{equation}
and the individual at-risk process
\begin{equation}
  \label{eq:5}
  Y_{i;l} (t) = \eins\left(X_i(t-)=l, L_i < t\le C_i\right)
\end{equation}
such that the counting process of observed $l\to m$ transitions is $N_{lm}(t)
= \sum_{i=1}^n N_{i;lm}(t)$ and the at-risk process for state~$l$ is $Y_{l}
(t) = \sum_{i=1}^n Y_{i;l} (t)$. Also introduce~$J_l(t)=\eins(Y_l(t)>0)$. We
assume that the~$N_{i;lm}$'s have absolutely continuous compensators with
respect to~$\mathcal{G}$, such that
\begin{equation}
  \label{eq:5komma5}
        M_{lm}(t) = N_{lm}(t) - \int_0^t \sum_{i=1}^n Y_{i;l} (u)\cdot \alpha_{i;lm}(u | \mathcal{G}(u-))\dif u,
\end{equation}
is a mean zero martingale with respect to~$\mathcal{G}$. If $(X(t))_{t\ge 0}$
is time-inhomo\-geneous Markov, the intensity $\alpha_{i;lm}(t |
\mathcal{G}(t-))$ will equal $\alpha_{lm}(t)$ from~\eqref{eq:3}, but in
general the intensity will be a random quantity through its dependence on the
past.

The Nelson-Aalen estimator is
\begin{displaymath}
  \hat{A}_{lm}(t) = \int_0^t \frac{J_l(u)}{Y_l(u)} \dif N_{lm}(u),
\end{displaymath}
and the Aalen-Johansen estimator is, using product integral notation,
\begin{equation}
  \label{eq:6}
  \prodint_{u \in (0,t]}
  \left(\mathbf{I}+\dif \hat{\mathbf{A}}(u)\right),
\end{equation}
where~$\mathbf{I}$ is the $(K+1)\times (K+1)$ identity matrix. The matrix
$\hat{\mathbf{A}}(t)$ has non-diagonal entries~$\hat{A}_{lm}(t)$ and diagonal
entries are such that the sum of each row equals zero.

The following result is similar to the classical strong consistency theorem of
the multivariate Nelson-Aalen estimator for time-inhomogeneous Markov
multi-state models \citep[][Theorem~IV.1.1]{ABGK}.
\begin{theorem}\label{theo1}
  Let $t \in [0,\tau]$ and $l,m \in \{0,1, \ldots, K\}, l\neq m$. Assume there
  exists a function~$k$, $\int_0^\tau k(u)\dif u < \infty$, such that
  \begin{equation}
    \label{eq:vor1}
    \alpha_{i;lm}(u | \mathcal{G}(u-)) \le  k(u) \mbox{\ on $[0,\tau]$ with probability~$1$},
  \end{equation}
  for all~$i=1, \ldots n$. Furthermore, as $n\to\infty$, assume that
  \begin{equation}
    \label{eq:vor2}
    \int_0^t \frac{J_l(u)}{Y_l(u)}k(u)\dif u \to 0 \mbox{\ in probability},
  \end{equation}
  and
  \begin{equation}
    \label{eq:vor3}
    \int_0^t (1 - J_l(u))k(u)\dif u \to 0 \mbox{\ in probability}.
  \end{equation}
  Then
  \begin{equation}
    \label{eq:kons1}
    \sup_{u\in [0,t]}\left|\hat A_{lm}(u) - A_{lm}(u)\right|\to 0 \mbox{\ in probability}.
  \end{equation}
\end{theorem}
The proof of Theorem~\ref{theo1} in the Appendix uses the proof of
Andersen et al. \citep[][Theorem~IV.1.1]{ABGK} as a template, but with the additional
difficulty that $\alpha_{i;lm}(t | \mathcal{G}(t-))$, $i=1,\ldots,n$, are
random quantities, unequal to $\alpha_{lm}(t)$. However, assuming i.i.d.\
multi-state trajectories, these random quantities are i.i.d., too, and their
average approaches $\alpha_{lm}(t)$.

Before we turn to estimating state occupation probabilities, some remarks on
Theorem~\ref{theo1} are in place:
\begin{enumerate}
\item In the time-inhomogeneous Markov case, the function~$k$ can be chosen as
  \begin{displaymath}
    k(t) := \sum_{l,m,l\neq m}\alpha_{lm}(t),
  \end{displaymath}
  because the transition hazards are assumed to have finite integrals
  \citep[][p.~287]{ABGK}.
\item In the presence of left-truncation, the convergence in probability
  statements are w.r.t.~the conditional probability measure given study
  entry~$Z$ from which we sample, see, e.g., Example~IV.1.7 of \cite{ABGK} and
  the Appendix. {\color{black}Also note that in the absence of left-truncation
    the main assumption both in our Theorem~\ref{theo1} and in the work by
    Datta and Satten and as compared to the classical result
    \citep[][Theorem~IV.1.1]{ABGK} on the Nelson-Aalen estimator is that
    right-censoring is entirely unrelated to the multi-state process.}
\item Analogously to the Markov case, a simple condition that implies
  assumptions~\eqref{eq:vor2} and~\eqref{eq:vor3} is that the infimum
  on~$[0,\tau]$ of all risk sets~$Y_l$ converges in probability to
  infinity. We refer to \cite{ABGK} for an in-depth discussion.
  This assumption may require reconsidering time~$0$ in left-truncated
  studies. E.g., in hospital epidemiology, a natural time origin is hospital
  admission. Studies with patients who are colonized with a certain infectious
  organism upon admission will typically include a substantial proportion of
  patients with colonization status known at time~$0$. Other patients will
  have left-truncated study entries upon arrival of laboratory results
  \citep[e.g.,\ ][]{de2011multistate}. In this setting, we may assume the
  condition to be fulfilled. However, in studies on pregnancy outcomes the
  natural time origin is conception, but women do not enter observational
  cohorts before pregnancy detection \citep[e.g.,\
  ][]{slama2014epidemiologic}. Time `zero' in the present context should then
  be chosen as the earliest time point of detecting pregnancies, around six
  weeks after the beginning of the menstrual cycle, or perhaps even slightly
  later, say 7 weeks.
\item In general, left-truncated data may contain information on
    the multi-state trajectory before study entry, but this information is not
    used here.
\end{enumerate}

Consistent estimation of the state occupation probabilities now follows from
Theorem~\ref{theo1}.
\begin{theorem}\label{theo2}
  Suppose~$\hat p(0)= (\hat p_1(0), \ldots, \hat p_K(0))$ is a consistent
  estimator of the initial distribution of the multi-state model,
  \begin{equation}
    \label{eq:t1}
    \hat p(0) \to \left(P(X(0)=1), \ldots, P(X(0)=K)\right) \mbox{\ in
      probability as $n\to\infty$},
  \end{equation}
  and define the $1\times K$ row vector
  \begin{equation}
    \label{eq:t2}
    \hat p(t) = \hat p(0) \cdot   \prodint_{u \in (0,t]}
    \left(\mathbf{I}+\dif \hat{\mathbf{A}}(u)\right).
  \end{equation}
  Then, under the assumptions of Theorem~\ref{theo1}, we have that
  \begin{equation}
    \label{eq:t3}
    \sup_{u\in [0,t]}\left|\hat p(u) - \left(P(X(u)=1), \ldots, P(X(u)=K)\right)\right|\to 0 \mbox{\ in probability}.
  \end{equation}
\end{theorem}

We prove Theorem~\ref{theo2} in the Appendix. The key assumption is
\eqref{eq:t1}. The choice of~$\hat p(0)$ is trivial, if there is one common
initial state, say state~$0$, with $P(X(0)=0)=1$. In the absence of
left-truncation, the obvious choice is the multinomial estimate $\hat p_j(0) =
Y_j(0+)/n$. With left-truncation, we will have to rely either on assuming one
common initial state, or on an educated `guess' $\hat p(0)$, perhaps from some
other data source, or on the fact that the individuals at risks are a random
draw from the underlying population, which leads to using $\hat p_j(0) =
Y_j(0+)/(\sum_l Y_l(0+))$. Interestingly, this difficulty
  disappears for the landmark estimator that we discuss next, because the
  landmark data set is constructed such that there is one common state
  occupied by all individuals at the landmark time.

\subsection{The landmark Aalen-Johansen estimator with
  left-truncation}\label{subsec:lmaj}

The landmark Aalen-Johansen estimator of
\cite{putter2016non} is based on subsampling as are the estimators of
\cite{de2015nonparametric} and \cite{titman2015transition}. The idea is
to select individuals that are {\color{black}under observation} in a given
state at a given time and estimate the state occupation probabilities within
this subset. Predating these contributions is the work by 
\cite{lida} who derived landmarking for this purpose in the special
illness-death model without recovery, already allowing, however, for delayed
study entry.

    To be precise, let
\[
  N_{lm}^{(\LM)}(t) = \sum_{i=1}^n N_{i;lm}(t)\mathbf{1}(L_i<s<C_i,\, X_i(s) = k),
  \quad s \leq t,
\]
be the counting process of the subsample that selects individuals
that are observed in state $k$ at time $s$. $N_{i;lm}(t)$ is defined as in
\eqref{eq:4}. Similarly, define
\[
  Y_{l}^{(\LM)}(t) = \sum_{i = 1}^n Y_{i;l}(t)\mathbf{1}(L_i<s<C_i,\, X_i(s) = k),
\]
be the subsample-based at-risk process \eqref{eq:5}. The landmark
Nelson-Aalen estimator is then
\[
\hat{A}_{lm}^{(\LM)}(t) = \int_0^t \frac{\eins(Y_l^{(\LM)}(u) > 0)}{Y_l^{(\LM)}(u)} \dif
  N^{(\LM)}_{lm}(u).
\]
Finally the landmark Aalen-Johansen estimator is given by
\[
  \hat P_{lm}^{(\LM)}(s, t) = \hat p^{(\LM)}(s) \cdot   \prodint_{u \in (0,t]}
  \left(\mathbf{I}+\dif \hat{\mathbf{A}}^{(\LM)}(u)\right)  \cdot (\hat p^{(\LM)}(s))^T,
\]
where $\hat p^{(\LM)}(s)$ is a row vector with entry~$1$ for
  state~$l$ and $0$ otherwise. 
Additionally assuming that $P(X(s) = l) > 0$, the
landmark Aalen-Johansen is a consistent estimator under the same assumption as
needed for the state occupation probabilities \citep{putter2016non}.

{\color{black}We emphasize that the landmarking approach, in general, uses less
  data than the standard Aalen-Johansen estimator. For illustration, consider
  an illness-death model without recovery, see also Section~\ref{sec:simu}
  below, but subject to right-censoring only. The Aalen-Johansen estimator of
  staying in the intermediate illness state on~$[s,t]$ given illness at
  time~$s$ is a standard Kaplan-Meier-type estimator. This estimator will also
  use observed trajectories entering the illness state after time~$s$, say, at
  time $\tilde s \in (s,t)$, and making a transition into the death state
  until time~$t$. The landmarking approach will not use such
  trajectories. Now, also introduce left-truncation. The standard
  Aalen-Johansen estimator would use, say, a trajectory that enters the study
  at time~$\tilde s$ in the illness state (and may even have been in the
  illness state at time~$s$). But the landmarking approach, now extended to
  left-truncated data, will not use this trajectory. The difference to the
  situation without left-truncation is that landmarking would have used this
  trajectory, if it had been in the illness state at time~$s$, but not, if it
  had fallen ill after time~$s$.}

{\color{black}For inference, we begin by exploiting} the i.i.d.\ structure of
the data under random left-truncation and right-censoring and use Efron's
bootstrap {\color{black} which draws with replacement from the units under
  study. To construct} point-wise confidence intervals, consider the
$(1 - \alpha/2)$-quantiles of the standardized bootstrap landmark
Aalen-Johansen estimator
\begin{equation}\label{eq:W_boot}
  W_n^* = \sqrt{n}\left(\hat{P}^{(\LM;*)}_{lm}(s, t) -
  \hat{P}^{(\LM)}_{lm}(s, t)\right)/\hat \sigma^{*},
\end{equation}
where $\hat \sigma^{*2}$ is the empirical variance of the bootstrapped
transition probability estimates, and plug them in the standard
asymptotic formula instead of the quantiles of the standard normal
distribution.

{\color{black}Alternatively, we re-express~\eqref{eq:5komma5} on the level of
  individual increments,
  \begin{displaymath}
    \dif M_{i;lm}(t) = \dif N_{i;lm}(t) - Y_{i;l} (t)\cdot \alpha_{i;lm}(t | \mathcal{G}(u-))\dif t,
  \end{displaymath}
  and substitute these unknown martingale increments by $\dif N_{i;lm}(t)$
  times a standard normal random variable as in \cite{bluhmkiWB}. Generating a
  large number of the latter multipliers given the data, i.e., treating
  $\dif N_{i;lm}(t)$ as fixed, is the basis of the wild bootstrap. A
  transformation of such simulated martingale distributions along the compact
  derivative of the product integral as in \cite{bluhmkiWB} yields another
  bootstrapping procedure. This wild bootstrap relies on an i.i.d.\ set-up in
  the present non-Markov setting subject to random left-truncation and
  right-censoring as does Efron's bootstrap. However, in a time-inhomogeneous
  Markov setting as in \cite{bluhmkiWB}, wild bootstrapping would also allow
  for more general censoring schemes, not necessarily entirely random and
  violating the i.i.d.\ structure, as we will demonstrate in
  Section~\ref{sec:wb}.}

\section{Simulation {\color{black}and real data} results}\label{sec:simu}

{\color{black} In both simulations and in the real data analysis, we focus on
  the illness-death model without recovery. The motivation from the real data
  analysis are hospital-acquired infections which will be represented by the
  intermediate `illness'-state. Departures from the Markov assumption occur,
  if the intensity of the illness-to-death transition also depends on the time
  of illness diagnosis. Section~\ref{sec:simul:so} uses simulations to study
  whether state occupation probabilities may be consistently estimated in a
  non-Markov model subject to random left-truncation. Section~\ref{sec:wb}
  takes a closer look at using Efron's bootstrap or the wild bootstrap. For
  ease of presentation, we consider the Nelson-Aalen estimator of the
  illness-to-death transition --- which `captures' violations of the Markov
  assumption --- and compare both bootstrapping procedures in a non-Markov
  setting and in a Markov setting. In the latter, censoring will not be
  random. Finally, simulations investigating the landmark Aalen-Johansen
  estimator are in Section~\ref{LAJ} and the real data example in
  Section~\ref{sec:data}.}

\subsection{State occupation probabilities}\label{sec:simul:so}

We present the results of a simulation study that assessed how well
the Aalen-Johansen estimator for the state occupation probability does
under random left-truncation. 

The simulations are based on a scenario used in
\cite{meira-machado06:_nonpar_markov}, who simulated an illness-death
model without recovery with initial state~0, intermediate state~1 and
absorbing state~2. Falling ill is modelled as a transition into
state~1, death is modelled as a transition into state~2 and recoveries
$1\rightarrow 0$ are not modelled. The hazards of a $0 \rightarrow 1$
and $0 \rightarrow 2$ transitions were assumed to be constant and
equal to 0.039 and 0.026, respectively.  The waiting time in the
initial state is generated using a constant hazard of 0.039 + 0.026. A
binomial experiment then decides on whether the individual moves into
state 1 with probability $0.039 / (0.039 + 0.026)$.

For the individuals that move into state 1, two methods for generating
times of arrival into state 2 are considered. The first simulation
method, suggested by \cite{couper}, is to specify $Z2 = (1 + d)Z1$,
where $d$ is an arbitrary constant and $Z1$ and $Z2$ denote the time
points of arrival in state 1 and 2, respectively. We set $d = 0.7$ in
the following \citep{lida}. The second simulation method uses a Cox
model to create the hazard function of a $1 \rightarrow 2$
transition. Let $\alpha_{12}(t|Z_1)$ be the hazard for a certain
individual to move from state 1 to state 2, with $\alpha_0$ the
baseline hazard and $\beta$ a constant coefficient. Then
$\alpha_{12}(t|Z_1)= \alpha_0 \exp (\beta Z_1)$.  The baseline hazard
for the $1 \rightarrow 2$-transition was set to 0.1 and the
coefficient $\beta=0.01$.

Random left-truncation times are sampled from a skew normal
distribution \citep{azzalini1985skew} with location, scale and shape
parameters chosen such that approximately 70\% of the individuals are
actually included in the study. Approximately 3\% of all simulated
individuals enter the study at time origin. The parameters are
$(0, 10, 10)$ and $(0, 13, 10)$ for the ``multiplication by a constant
scenario'' and the ``Cox scenario'', respectively.

\begin{figure}[t]
  \centering
  \includegraphics[width = \textwidth]{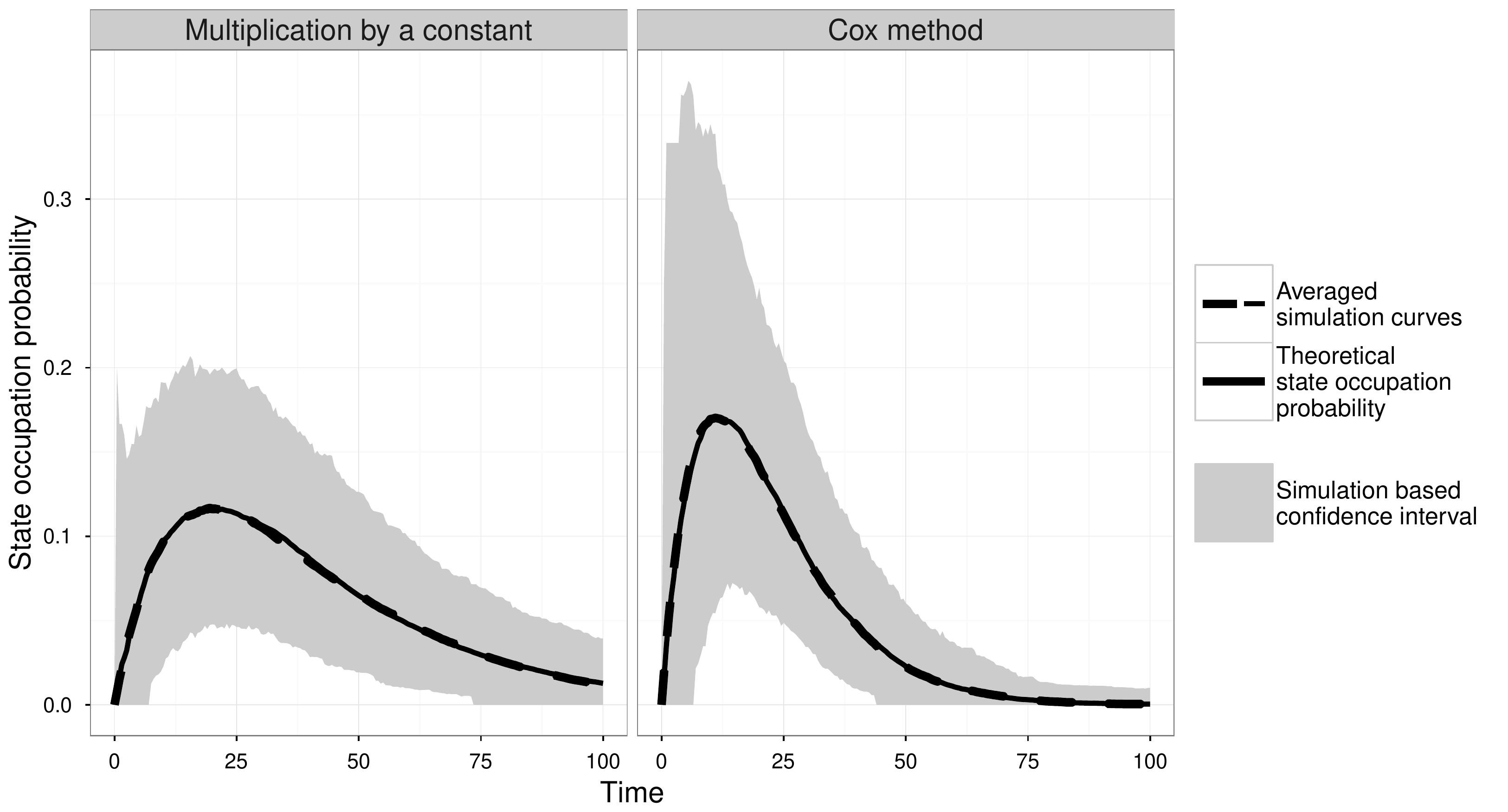}
  \caption{Results of the simulation study for the scenario
    ``multiplication by a constant'' in the left panel and the ``Cox
    method'' on the right. The solid black line represents the true
    state occupation probability, the dashed line the averaged
    probabilities obtained from the simulated studies. The grey area is
    the simulation based 95\% confidence interval.}
  \label{fig:simulation}
\end{figure}

We simulated 1000 studies with a sample size of 100
individuals. Figure~\ref{fig:simulation} reports the average of the
1000 estimates of $P_{01}(0, t)$ as well as the simulation based 95\%
confidence intervals for the two scenarios. Also displayed is the true
value $P_{01}(0, t)$ numerically approximated by computing the
Aalen-Johansen estimator in a study without truncation with 100.000
individuals.

As can be seen, the curves can almost not be distinguished. Therefore,
we can conclude that within these simulation designs the
Aalen-Johansen estimator for the state occupation probability still
performs well under random left-truncation. 

\subsection{The wild bootstrap resampling technique}\label{sec:wb}
{\color{black}
  Before we investigate more closely the performance of the landmark Aalen-Johansen estimator in the next section, we {\color{black}consider} in this section the wild bootstrap resampling technique in non-Markov models and compare it with the standard Efron's bootstrap. Moreover, to get a more complete picture of the performance of the wild bootstrap, we extend our evaluation to Markov models subject to type II censoring.    \\
  The key result, that allows us to apply the wild bootstrap in non-Markov
  models, is the consistent estimation of the cumulative transition hazard by
  the Nelson-Aalen estimator as shown in Theorem~\ref{theo1}. Therefore, we
  focus in this section on the Nelson-Aalen estimator of the cumulative hazard of the 1 $\rightarrow$ 2 transition. Detailed information to
  the wild bootstrapping of the multivariate Nelson-Aalen estimator including
  the transformation onto the Aalen-Johansen estimator can be found in
  \cite{bluhmkiWB}. \\As a first step, we simulated data from an illness-death
  model without recovery as in the previous section. We introduce dependence
  between the waiting time in the initial state and the waiting time in the
  illness state by multiplying constant transition hazards by a common gamma
  frailty Z. Here, Z is a gamma-distributed random variable with mean and
  variance equal to 2. Moreover, we added exponentially distributed random
  right-censoring times. We use the following constant transition hazards:
  $\alpha_{01}=0.12 $, $\alpha_{12}=0.1 $, $\alpha_{02}=0.03 $.  We considered
  different sample sizes --- 30, 50 and 100 individuals per study --- and
  simulated 100 studies for each scenario.  For the construction of point-wise
  95\% confidence intervals, we used two different resampling techniques ---
  Efron's bootstrap and the wild bootstrap technique. Both resampling methods
  are used to determine the $(1- \alpha/2)$ - quantiles of the standardized
  Nelson-Aalen estimator which were plugged-in in the standard asymptotic
  formula instead of using the quantiles of the standard normal
  distribution. We use the empirical variance of the bootstrapped Nelson-Aalen
  estimators as variance estimator. Table \ref{tab:NA} compares the coverage
  probabilities of the 95\% point-wise confidence intervals obtained from
  those two resampling methods for different sample sizes and at different
  time-points. Both methods provide coverage probabilities close to the
  nominal level of 95\%. However, for the scenarios with a sample size of 30
  individuals, the wild bootstrap approach still leads to coverage
  probabilities close to the nominal level, whereas the confidence intervals
  obtained by Efron's bootstrap are quite liberal.  In summary, under an
  i.i.d.\ data structure both resampling approaches {\color{black}provide
    reliable results} in situations where the Markov property is in doubt. Our
  simulations indicate that for small sample sizes the wild bootstrap approach
  performs better compared to Efron's bootstrap. \\ One big advantage of the
  wild bootstrap technique is, that it is, in contrast to Efron's bootstrap,
  not limited to the strict situation with i.i.d.\ data structure. Thus, the
  wild bootstrap does not require random censoring. As the i.i.d.\ data
  structure is a requirement for the consistency of the Nelson-Aalen estimator
  in non-Markov models, we consider a Markov model subject to event-driven
  censoring, so-called type II censoring, to investigate the impact of the
  violation of the i.i.d.\ data assumption. {\color{black}In other words, the
    aim of the following simulation is to investigate possible advantages of
    wild bootstrapping when random censoring, but not the Markov property is
    in doubt.}  Type II censoring implies that all individuals will be
  censored at the time when a specified number of occurrences of the event of
  interest has been taken place. Thus, type II censoring is no random
  censoring but it fulfills the independent censoring assumption
  {\color{black}of \cite{ABGK} and} \cite{aalen2008survival}{\color{black}, in
    that retains the form of the intensities of the counting processes as in
    \eqref{eq:5komma5}.}

In our simulation studies, we censored all individuals at the time when half of the individuals had an observed death event. We used constant hazard rates ($\alpha_{01}=0.01 $, $\alpha_{12}=0.1 $, $\alpha_{02}=0.03 $) and no staggered study entry. That means all individuals enter the study at time 0. Table \ref{tab:NA_2} shows the coverage probabilities of the 95 \% confidence intervals constructed using the two different resampling methods at different time points and for different sample sizes. It can be seen that the wild bootstrap technique provides coverage probabilities closer to the nominal level compared to Efron's bootstrap for all considered scenarios.

\begin{table}[ht]
 \centering
  \caption{Comparison of empirical coverage probabilities of the 95\% point-wise confidence intervals constructed using Efron's bootstrap and the wild bootstrap at different time-points in non-Markov illness death model}\label{tab:NA}
\begin{tabular}{l rrr@{\hspace{1cm}}rrr}
\toprule
 & \multicolumn{6}{c}{Coverage (\%)} \\
 &  \multicolumn{3}{c}{Efron} & \multicolumn{3}{c}{Wild Bootstrap}\\                    
 n & T15 & T20 & T25  & T15 & T20 & T25  \\             
\midrule
n=30 & 88 &87 & 81 &96 &97 &97\\
n=50 & 94 &93  &91  &97  &96   &95\\
n=100  &98 &98 & 97 &  95 & 97  & 98\\

\bottomrule
\end{tabular}
\end{table}

\begin{table}[ht]
 \centering
 \caption{Comparison of empirical coverage probabilities of the 95\%
   point-wise confidence intervals constructed using Efron's bootstrap and
   wild bootstrap at different time-points in Markov models subject to type II
   censoring {\color{black}with censoring after~$m$ events have been
     observed.}}\label{tab:NA_2}
\begin{tabular}{ll rrr@{\hspace{1cm}}rrr}
\toprule
 &  &\multicolumn{6}{c}{Coverage (\%)} \\
 & & \multicolumn{3}{c}{Efron} & \multicolumn{3}{c}{Wild Bootstrap}\\                    
 n & m & T8 & T12 & T16  & T8 & T12 & T16  \\             
\midrule

n=80 &m=40  &71 &83 & 87 &  88 & 98  & 97\\
n=100& m=50 &81 &87 & 87 &  94 & 97  & 97\\
n=200& m=100 &88 &91 &93  &93   &92  &95 \\
\bottomrule
\end{tabular}
\end{table}

}

\subsection{The landmark Aalen-Johansen estimator}\label{LAJ}

We now extend the simulation setting of \cite{titman2015transition} and
\cite{putter2016non}. The data are simulated from an
illness-death model without recovery. As in Titman, we consider two processes,
termed `Frailty' and `non-Markov'. The baseline transition hazards are
constant, with $\alpha_{01} = 0.12$, $\alpha_{02} = 0.03$ and $\alpha_{12} =
0.1$. For the 'Frailty' model, all hazards are multiplied by a common gamma
frailty $Z$ with mean and variance equal to 2. The frailty term
  introduces dependence between the waiting time until leaving the initial
  state of the illness-death model and the waiting time until absorption and,
  hence, a violation of the Markov assumption. In the 'non-Markov' scenario,
$\alpha_{12}(t)$ is dependent on the state occupied at time 4, i.e.,
\[
  \alpha_{12}(t) =
  \begin{cases}
    0.05,& \text{if } X(4) = 0,\\
    0.1, & \text{otherwise}.
  \end{cases}
\]
Transition times were simulated as in Section~\ref{sec:simul:so}. We
considered sample sizes $m=200$ and $m=500$. Then random left-truncation times
following a Uniform distribution ({\em Unif}) with parameters {\color{black} $(-5, 28)$ and
$(-1, 13)$ for the `Frailty' and `non-Markov' scenario, respectively. We consider also
exponentially distributed left-truncation times with parameter
$0.13$. Here, nobody is starting in the study
at time 0.} Table~\ref{tab:simul} reports number of individuals
  simulated ($m$), average number of individuals in the study
($\bar{n}$, $\bar{n}<m$ because of left-truncation), bias, root
mean squared error (RMSE), and {\color{black} the empirical coverage probability of the 95\% bootstrap quantile
confidence interval \eqref{eq:W_boot} ({\em Cov})}, for the Aalen-Johansen
and landmark Aalen-Johansen estimates of $P_{01}(\tau_{0.15}, \tau_{0.45})$,
where $\tau_{0.15}$ and $\tau_{0.45}$ correspond to the 15th and 45th
percentiles of the time-to-absorption distribution whose values are taken from
the supplementary material of \cite{titman2015transition}.

\begin{table}
  \centering
  \caption{Average number of individuals under study ($\bar{n}$),
    bias, root mean square error (RMSE), and coverage}\label{tab:simul}
\begin{tabular}{lrrrrrrlrrl}
  \toprule
    &     &   & \multicolumn{3}{c}{AJ} & \multicolumn{3}{c}{LMAJ}                               \\
\cmidrule{4-6}
\cmidrule{7-9}
Trunc & $m$ & $\bar{n}$ & Bias & RMSE  & Cov (\%)   & Bias   & RMSE  & Cov (\%)  \\ 
  \midrule
{}\\
\multicolumn{9}{c}{Simulation model `Frailty'}\\

Unif  & 200 &   141   &  -0.0038    &  0.047 & 97 &  0.0001 & 0.069   & 92   \\ 
      & 500 &   353   &   0.0003    &  0.028 & 97 &  0.0056 & 0.038  & 99   \\ 
      
Exp   & 200 &   157   &  -0.0053    &  0.040 & 91 & 0.0015  & 0.051 & 96 \\ 
      & 500 &   391   &  -0.0016    &  0.023 & 95 & 0.0008  & 0.032  & 97  \\
            
{}\\
\multicolumn{9}{c}{Simulation model `non-Markov'}\\
Unif  & 200 & 152    &  -0.0250    & 0.060 & 93  &  0.0087 & 0.083 & 97 \\ 
      & 500 & 382    &  -0.0285    & 0.041 & 85  & -0.0010 & 0.046 & 98  \\ 
Exp   & 200 & 146    &  -0.0266    & 0.057 & 88  & -0.0008 & 0.081 & 92 \\ 
      & 500 & 362    &  -0.0280    & 0.043 & 89  & -0.0040 & 0.048 & 93 \\
   \bottomrule
\end{tabular}
\end{table}

As in \cite{putter2016non}, the landmark Aalen-Johansen performs
well. The Aalen-Johansen estimator is slightly more biased but displays the smallest
RMSE for most scenarios. Efron's bootstrap provides confidence intervals with coverage probabilities close to the nominal level for the Aalen-Johansen estimator in the `Frailty' model, whereas in the `non-Markov' model the  coverages of that estimator are more liberal. With regard to the landmark Aalen-Johansen estimator, the coverages are similar in both models.  
 
An alternative to Efron's bootstrap is the wild bootstrap resampling technique. As pointed out in Section \ref{sec:wb}, this approach is valid in non-Markov models and can be used to construct confidence intervals for the Nelson-Aalen estimator. Following the proceeding of \cite{bluhmkiWB}, we assume that the wild bootstrap can be also applied for construction of confidence intervals for the landmark Aalen-Johansen estimator, but this is subject to further research.

\subsection{Real data example: {\color{black}n}osocomial infection and stay in
  {\color{black}h}ospital}\label{sec:data}

We consider data {\color{black}on patients colonised by} Methicillin-resistant
{\em staphylococcus aureus} (MRSA) {\color{black}from } a prospective cohort
study in 12 surgical units at the University of Geneva Hospitals, Switzerland,
between July 2004 and May 2006 \citep{de2011multistate}. {\color{black}MRSA
  carriage is not necessarily detected upon hospital admission, because a
  positive MRSA screening result may come in `delayed' in the sense that the
  positive laboratory result becomes available or known only after
  admission. Hence, our population of interest are patients colonised by MRSA,
  the time scale of interest is time since hospital admission and the
  left-truncation time is the time of detecting MRSA in the screening
  process. Colonised patients who are discharged or die before a positive
  screening result becomes available are not included in the study.

  MRSA colonization may lead to MRSA \emph{infection} in hospital, and the}
more severe or potentially life-threatening MRSA infections are observed most
frequently in healthcare settings. {\color{black}In this context and in the
  presence of limited financial resources, possible measures of infection
  control are weighed against the costs associated with hospital-acquired
  infection \citep{muto2003shea}. To this end, excess length of stay following
  the infection is typically considered to be the main cost driver
  \citep{graves:2010}.

  However, quantifying excess length of stay is complicated by the fact that
  hospital-acquired MRSA infection is a time-dependent exposure and a simple
  retrospective comparison of the distribution of length of stay of the
  infected with that of the only colonised must overestimate the prolonging
  effect of the infection as a consequence of `immortal time bias'
  \citep{beyersmann07_iscb07,suissa08}. We address this difficulty as follows:
  Firstly, we model occurrence of hospital-acquired MRSA infection as an
  intermediate state} in an illness-death {\color{black}multi-state} model, in
which the initial state represents colonization, {\color{black}intermediate}
state 1 infection and {\color{black}absorbing} state 2 discharge from the
hospital. {\color{black}Secondly, we use landmarking to compare the residual
  length of stay (in terms of the transition probabilities) of those in
  in the infectious state at the landmark with those still in the initial
  state of colonization.

  Two remarks are in place: Firstly, an alternative modelling approach would
  be a cure model, where a `cure proportion' of colonised patients `immune' to
  infection accounts for the fact that only a fraction of the colonised
  patients are diagnosed with MRSA infection in hospital. This is in contrast
  to the multi-state approach where the interplay of the intensities out of
  the initial colonization state, one for infection, one for direct discharge,
  regulates the proportion of infected patients. One reason to choose a
  multi-state modelling approach was that not becoming infected may be a
  consequence of actions taken after a positive screening result such as
  decolonization \citep{de2011multistate}. Secondly, landmarking has been
  introduced in medical research to deal with the difficulties of comparing
  groups defined by time-dependent exposures
  \citep{Ande:Cain:Gelb:anal:1983,JRA2008}, while the landmark Aalen-Johansen
  estimator of  \cite{putter2016non} and of our
  Section~\ref{LAJ} has used landmarking for estimating transition
  probabilities rather than just state occupation probabilities. This further
  highlights the close connection between time-dependent exposures and
  multi-state modelling.
}


{\color{black}We begin our analysis by checking the Markov assumption using a
  Cox model, estimate the proportion of currently infected and hospitalized
  patients using the Aalen-Johansen estimator of state occupation
  probabilities and finally compare the residual length-of-stay distributions
  between infected and only colonised for different landmarks using the
  landmark Aalen-Johansen estimator of transition probabilities. Recall that
  all analyses must account for the data being subject to left-truncation.

  In order to} check the Markov assumption {\color{black}we include} the time
of infection in a Cox proportional hazards model for the hazard of
end-of-stay~\citep{Keid:Gill:rand:1990}. The HR is found to be significantly
smaller than 1 (HR: 0.98, 95\%-CI [0.97, 0.99]). Thus the later one becomes
infected the lower the hazard of being discharged.

\begin{figure}[t]
  \centering
  \includegraphics[width = \textwidth]{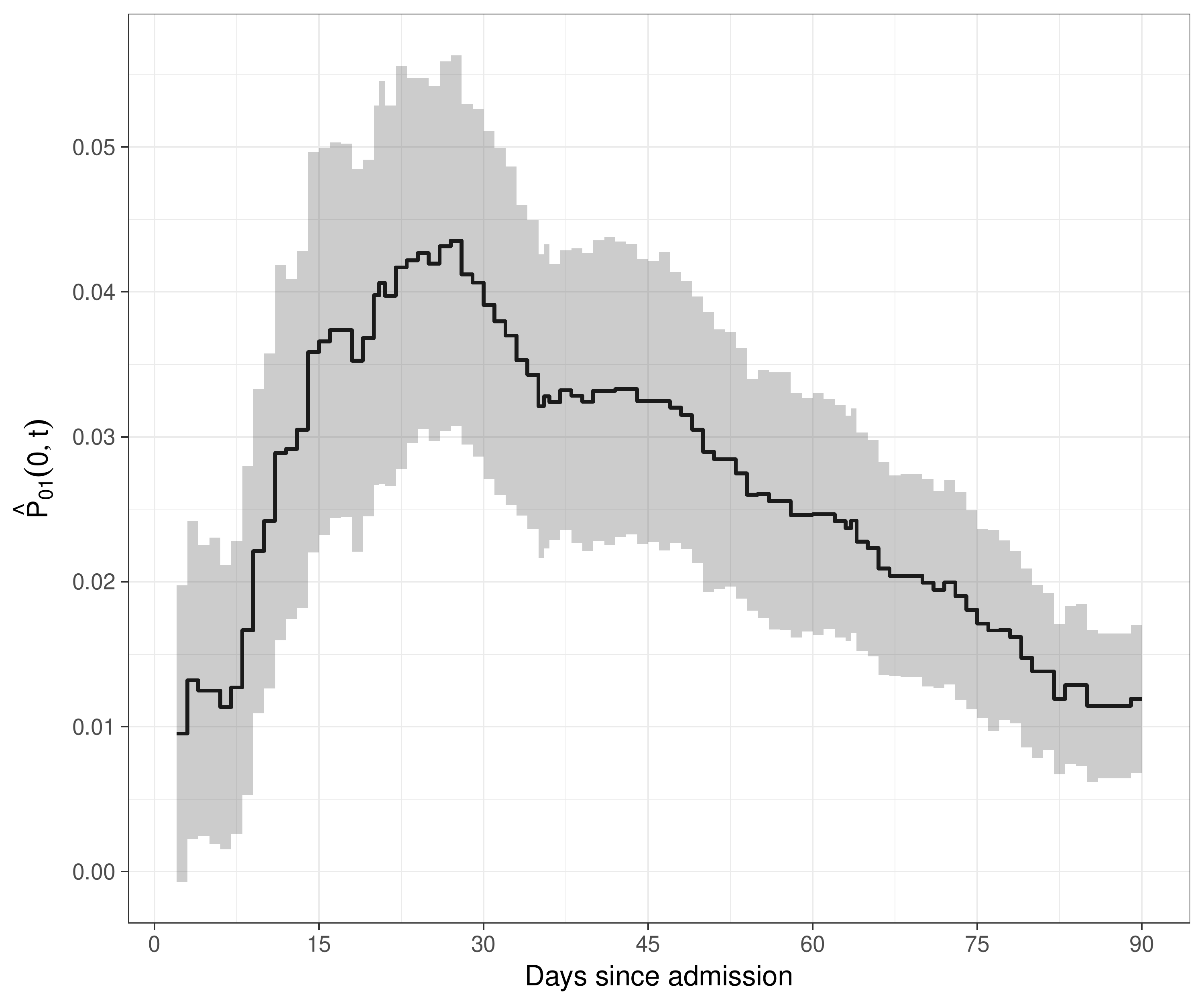}
  \caption{{\color{black}Estimated s}tate occupation probability $P_{01}(0, t)$
    for the MRSA data along with 95\% point-wise confidence interval based on
    1000 bootstrap samples.}\label{fig:p01}
\end{figure}
Figure \ref{fig:p01} displays the {\color{black}estimated} probability to be
{\color{black}in the infectious state}, i.e., the state occupation probability
based on 1000 bootstrap samples. We note that this probability is low.

\begin{figure}[t]
  \centering
  \includegraphics[width = \textwidth]{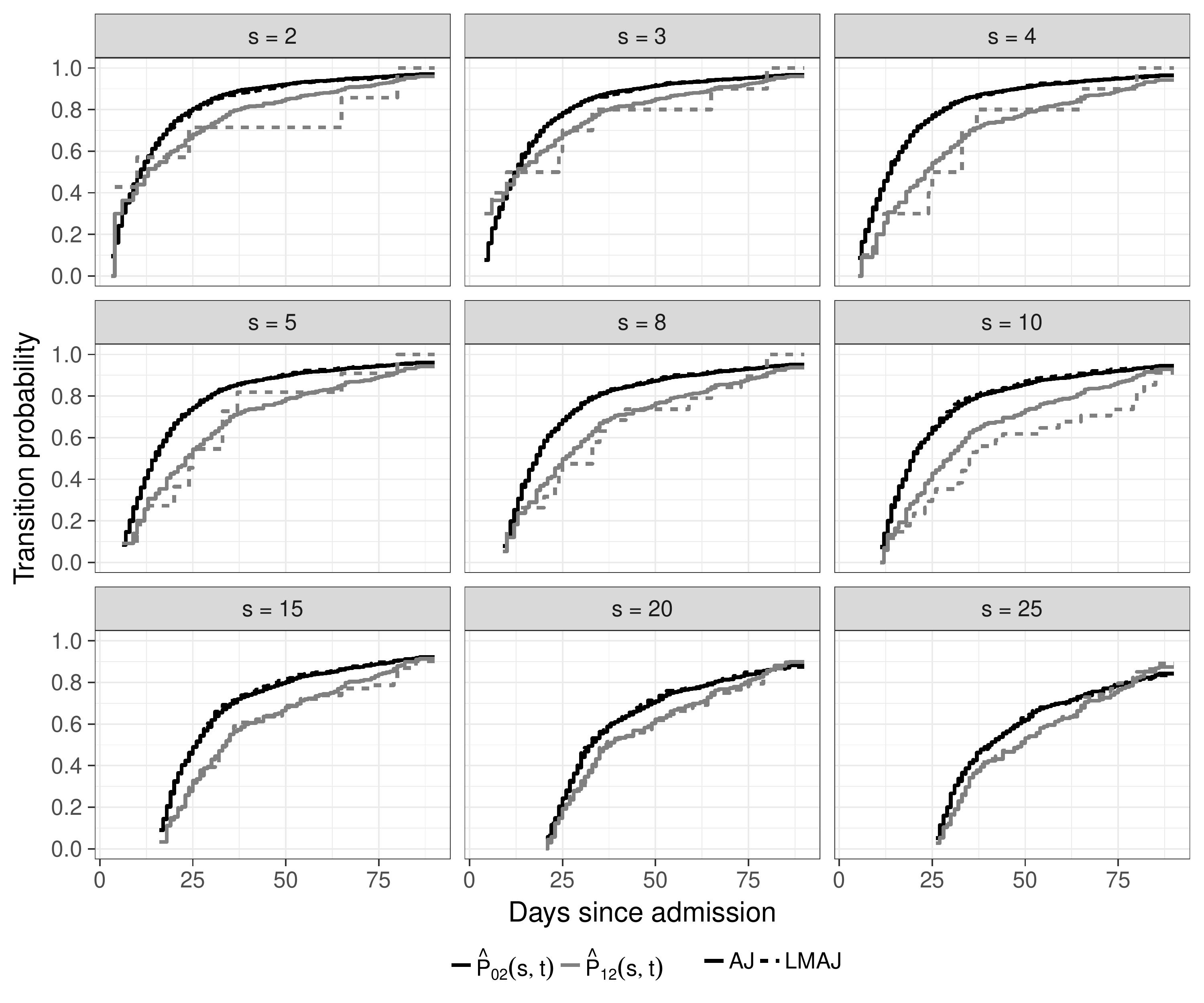}
  \caption{Landmark analysis of the MRSA data. Each panel displays
    $P_{02}(s, t)$ (black) and $P_{12}(s, t)$ (grey) for
    $s \in \{2, 3, 4, 5, 8, 10, 15, 20, 25\}$ estimated by the
    Aalen-Johansen estimator (solid lines) and the landmark
    Aalen-Johansen estimator (dashed lines).}\label{fig:landmark}
\end{figure}


{\color{black}Finally,} Figure~\ref{fig:landmark} provides a landmark analysis,
display{\color{black}ing both landmark Aalen-Johansen estimates and the
  original Aalen-Johansen estimates of} $P_{02}(s, t)$ and $P_{12}(s, t)$ for
a selected range of landmark times\\ $s = \in \{2, 3, 4, 5, 8, 10, 15, 20, 25\},
\, s < t$. The prolonging effect of MRSA is illustrated by the fact that
$\hat{P}_{02}(s, t)$ runs above $\hat{P}_{12}(s, t)$ for all $s$, though this
effect is much more pronounced for s between 8 and 15 days. We also see that
for these data, the Aalen-Johansen and landmark Aalen-Johansen estimators are
close to each other in spite of the data being possibly non-Markov.

\section{Discussion}\label{sec:disc}

Multi-state modelling is useful for investigating complex courses of a disease
in a variety of medical disciplines, but usually come{\color{black}s} with a
time-inhomogeneous Markov assumption to facilitate the technical
developments. In particular, estimating transition probabilities when the
Markov assumption is in doubt has been an active field of research in recent
years. To the best of our knowledge, one of the first proposals is due to
\cite{meira-machado06:_nonpar_markov} who used Kaplan-Meier integrals for a
randomly right-censored illness-death model. Their approach was simplified by
\cite{allignol2013competing} using competing risks techniques. Allignol et
al.\ also used semi-parametric efficiency arguments to arrive at landmark
transition probability estimators, also allowing for delayed study entry, see
also \cite{marcimarc}. Their approach was then recently extended to arbitrary
multi-state models by \cite{titman2015transition}, see also
\cite{de2015nonparametric} for related work. Arguably the most natural
approach is due to \cite{putter2016non} using a landmark Aalen-Johansen
estimator and consistency of the Aalen-Johansen estimator for state occupation
probabilities of non-Markov multi-state models subject to random
right-censoring. Our paper complements the work by Putter and Spitoni,
establishing the consistency needed and extending results to delayed study
entry which is a common phenomenon in observational studies. {\color{black}Our
  proof also motivates use of } {\color{black} the wild bootstrap resampling
  method and shows its validity in non-Markov models using simulation
  studies. In contrast to Efron's bootstrap the wild bootstrap is not limited
  to situations with i.i.d.\ data structure and, hence, could also be applied
  to censoring scenarios {\color{black}that are more complex than } random
  censoring \citep{bluhmkiWB}}, {\color{black}then, however, relying on the
  time-inhomogeneous Markov framework}.

\paragraph{Acknowledgement} Jan Beyersmann was {\color{black}partially}
supported by Grant BE 4500/1-2 of the German Research Foundation (DFG).

\bibliographystyle{wileyauy}
\bibliography{literatur}

\appendix
\section{Proofs}

\begin{proof}
  of Theorem~\ref{theo1}

  To begin, note that we do not sample from~$P$ but from the conditional
  probability measure given study entry~$Q(\cdot) = P(\cdot\,|\,Z)$.  Dropping
  indices~$l$ as in $Y_{i;l}$ and~$lm$ as in $N_{lm}$ denoting the transition
  type for ease of notation, we have that
  \begin{eqnarray*}
    \hat{A}(t) & = & \int_0^t \frac{J(u)}{Y(u)} \dif N(u)\\
    {} & = & \underbrace{\int_0^t \frac{J(u)}{Y(u)} \dif M(u)}_{=M^\star(t)} + \underbrace{\int_0^t \frac{J(u)}{Y(u)}
    \sum_{i=1}^n Y_{i} (u)\cdot \alpha_i(u | \mathcal{G}(u-))\dif u}_{=A^\star(t)}.
  \end{eqnarray*}
  $M^\star(t)$ is a mean zero martingale with predictable variation process
  \begin{displaymath}
    V(t) = \int_0^t \frac{J(u)}{Y(u)^2}\sum_{i=1}^n Y_{i} (t)\cdot \alpha_i(u | \mathcal{G}(u-))\dif u.
  \end{displaymath}
  Because of Lenglart's inequality, we have that for any~$\delta,\eta>0$
  \begin{displaymath}
    Q\left(\sup_{[0,t]}|\hat A - A^\star|>\eta\right) \le
    \frac{\delta}{\eta^2} + Q(V(t)>\delta).
  \end{displaymath}
  Assumption~\eqref{eq:vor1} implies
  \begin{displaymath}
    V(t) \le \int_0^t \frac{J(u)}{Y(u)}k(u)\dif u
  \end{displaymath}
  and it follows from~\eqref{eq:vor2} that
  \begin{displaymath}
    \sup_{[0,t]}|\hat A - A^\star| \stackrel{Q}{\to} 0.
  \end{displaymath}
  To complete the proof, we still need to show
  \begin{equation}
    \label{eq:show}
    \sup_{[0,t]}\left|\int_0^t \left\{\frac{J(u)}{Y(u)}
        \sum_{i=1}^n Y_{i} (u)\cdot \alpha_i(u | \mathcal{G}(u-))\right\} - \alpha(u)\dif
      u\right| \stackrel{Q}{\to} 0.
  \end{equation}
  Under assumption~\eqref{eq:vor1} and using Gill's dominated convergence
  theorem \citep[][Proposition~II.5.3]{ABGK}, it suffices to show point-wise
  convergence in probability of the integrand. We have
  \begin{displaymath}
    \frac{Y(u)}{n}\stackrel{Q}{\to} Q(Y_1(u)=1) = Q(Y_i(u)=1)\mbox{\ for all\ }i=1,\ldots, n.
  \end{displaymath}
  Next, note that the dependence of $\alpha_i(u | \mathcal{G}(u-))$ on
  the past of the observed data only constitutes dependence on $i$th observed
  past but not on that of other individuals~$j$, $j\neq i$, because of
  independence across individuals. Using that left-truncation and
  right-censoring are random, we get that
  \begin{displaymath}
    Y_{i} (u)\cdot \alpha_i(u | \mathcal{G}(u-)), i=1,\ldots, n,
  \end{displaymath}
  are i.i.d.\ random variables, and their average approaches the mean given by
  the following calculation
  \begin{eqnarray*}
    \erw(Y_{i} (u)\cdot \alpha_i(u | \mathcal{G}(u-))\dif u) & = &
    \erw\big(\erw\{Y_{i}(u)\cdot \alpha_i(u | \mathcal{G}(u-))\dif u\}\,|\,
    \mathcal{G}(u-)\big)\\
    {} & = & \erw\big(Q({\rm d} N_i(u)=1\,|\, \mathcal{G}(u-))\big)\\
    {} & = & Q({\rm d} N_i(u)=1)\\
     {} & = & Q(Y_i(u)=1)\cdot  Q({\rm d} N_i(u)=1\,|\,Y_i(u)=1)
  \end{eqnarray*}
  and, recalling that we consider transitions~$l\to m$,
  \begin{eqnarray*}
    \lefteqn{Q({\rm d} N_i(u)=1\,|\,Y_i(u)=1)}\\
    {} & = & P(X_i(u+{\rm d}u)=m, u+{\rm d}u\le
    C\,|\, X_i(u-)=l, L_i<u\le C_i)\\
    {} & = &P(X_i(u+{\rm d}u)=m\,|\, X_i(u-)=l) = \alpha(u)\dif u.\\
  \end{eqnarray*}
  In the previous display, the first equality holds, because $X_i(u-)=l,
  L_i<u$ implies study entry for a transient state~$l$. Point-wise convergence
  of the integrand in~\eqref{eq:show} follows, which completes the proof.
\end{proof}

\begin{proof}
  of Theorem~\ref{theo2}

  The proof relies on the observation by Andersen et
  al. \citep[][Section~IV.4.1.4]{ABGK} for complete data that the
  entries of $\hat p(t)$ are the usual multinomial estimates, i.e.,
  the number of trajectories observed in a specific state divided
  by~$n$, if we chose $\hat p(0)$ as remarked after
  Theorem~\ref{theo2}. In other words, Theorem~\ref{theo2} holds in
  the absence of both left-truncation and right-censoring.

  Now, because product integration is a continuous functional {\color{black}(or
    operator)} \citep{Gill:Joha:surv:1990}, the assertion will follow as a
  consequence of the continuous mapping theorem, if the Nelson-Aalen estimator
  consistently estimates the same limit in the presence of random
  left-truncation and right-censoring as it does in the complete data
  case. This is precisely what Theorem~1 states, which completes the
  proof. {\color{black}Here, we view the multivariate Nelson-Aalen estimator as
    a random element of $D[0, t]^{K^2 + K}$, $t\le\tau$, equipped with the
    max-supremum norm.}
\end{proof}

\end{document}